\documentclass[reprint,3p,twocolumn,sort&compress]{elsarticle}
\usepackage{amsfonts} 
\usepackage{amsmath}
\usepackage{amssymb}
\usepackage{graphicx}
\usepackage{subfigure}
\usepackage{color}
\journal{Phys. Lett. A}
\begin{document}
\begin{frontmatter}
\title{Surface plasmons  in a semi-bounded  massless Dirac plasma }
\author[MS]{M. Shahmansouri\corref{cor1}}
\ead{mshmansouri@gmail.com}
\cortext[cor1]{Corresponding author}
\author[MS]{R. Aboltaman}
\address[MS]{Department of Physics, Faculty of Science, Arak University, Arak 38156-8 8349, Iran}
\author[APM]{A. P. Misra}
\ead{apmisra@visva-bharati.ac.in; apmisra@gmail.com}
\address[APM]{Department of Mathematics, Siksha Bhavana, Visva-Bharati University, Santiniketan-731 235, India}
 
\begin{abstract}
The collective excitation   of   surface plasmons in a massless Dirac plasma (e.g., graphene) half-space (bounded by air)  is investigated using a relativistic quantum fluid model. The unique features of such surface waves are discussed and compared with those in a Fermi plasma. It is found that in contrast to Fermi plasmas, the long-wavelength surface plasmon frequency $(\omega)$   in massless Dirac plasmas is explicitly  nonclassical, i.e.,   $\omega\propto1/\sqrt{\hbar}$, where $h=2\pi\hbar$  is the Planck's constant. Besides some apparent  similarities between the surface plasmon frequencies in massless Dirac plasmas and Fermi plasmas, several notable differences are also found and discussed. Our findings elucidate the properties of surface plasmons that may propagate in degenerate plasmas   where  the relativistic and quantum effects  play a vital role.
\end{abstract}
\end{frontmatter}
\section{Introduction}\label{intro}	  
The collective oscillations of interacting electrons (i.e., plasmons), have attracted a considerable attention due to their potential applications, e.g.,  in exploring the effects of electron-electron   interactions   in different physical systems  including optical metamaterials, in receiving light signals at the nanoscale, in ultrafast lasers, in solar cells, in photodetectors, in biochemical sensing, as well as, in  transmitting antennas \cite{pines1966,giuliani2005,bonaccorso2010,geim2011,novoselov2004,grigorenko2012,fei2013,
li2008,neto2009,liu2014,bonaccorso2012,peres2010}. A number of theoretical works  \cite{pines1966,giuliani2005,ando1982} on collective modes of ordinary (Schr{\"o}dinger) electrons in Fermi plasmas   and their experimental verifications \cite{ando1982,mahan2000,li1991} are already in the  literature.   The classical plasma frequency in three-dimensional (3D) plasmas   is known to be $\omega_p=\sqrt{4\pi n_0e^2/m}$, where $n_0$ is the unperturbed number density and $m$   the mass of electrons. Though  this frequency appears in Fermi plasma fluids, it may not be the same in massless Dirac plasmas, such as those in, e.g., graphene.   
\par
Because of its peculiar features and amazing electronic and optical properties,  graphene has attracted a huge interest in recent years.  The dense honeycomb arrangements of carbon atoms with photon-like massless energy relation  have made it possible for the charge carriers in graphene to mimic both relativistic and quantum effects at the same time \cite{moghanjough2013a}. Such massless electrons can move with an effective Fermi speed  of about $v_F\sim10^6$ m/s,   which is independent of the carrier number density. It has been shown that the dynamics of two-dimensional (2D) gas of charged particles in graphene   can be described by the relativistic Dirac fluid model \cite{sarma2009,sachdeva2015}. In this context, the linear-band dispersion of Dirac electrons in graphene is known to be the origin of some new features in wave dynamics that are distinctive from  the ordinary 2D degenerate electron gas \cite{sarma2009}.
\par
Dirac materials (particularly in graphene) \cite{grigorenko2012,ju2011,yan2012,yoon2014,yan2013,stauber2014}  have been considered for the excitation of plasmons due to their tunable spectrum through the electrostatic control of their carrier concentration, and also their high lifetime plasmons (because of high mobility). A number of authors have proposed the theory of plasmons in Dirac systems in various forms, such as topological insulators \cite{pietro2013,raghu2010}, graphene \cite{sarma2009,wunsch2006,hwang2007,polini2008,abedinpour2011}, Weyl semi-metals \cite{zhou2015}, graphene microribon arrays \cite{ju2011}, and massless Dirac plasma layers \cite{sarma2009,zhu2013,triola2012}.
\par
The propagation of electrostatic surface waves in semi-bounded plasmas have been studied by Ritchie \cite{ritchie1963} and  the effects of  finite  temperature on these surface waves have also been discussed by using a hydrodynamic model.  The theory of Ritchie was later extended to a quantum plasma half-space using a quantum hydrodynamic (QHD) model by Lazar \textit{et. al.} \cite{lazar2007}.  Furthermore, the dispersion properties of surface Langmuir oscillations have been studied by Chang \textit{et. al.} \cite{chang2008} in a semi-bounded quantum plasma  using the specular reflection method. 
 Such QHD model has been known to be one of the powerful models for the investigation of wave dynamics   in quantum plasmas \cite{kaw1970,misra2011,misra2010,niknam2013,shahmansouri2015,moradi2015,moradi2016a,
shahmansouri2017a,moradi2017,shahmansouri2017b,moghadam2017,shahmansouri2017c,tyshetskiy2013,
zhu2013b,moradi2016b,misra2007}. It has been  shown that the propagation characteristics of surface waves can be modified by the effects of quantum tunneling \cite{kaw1970,misra2011,misra2010,niknam2013,shahmansouri2015,moradi2015,moradi2016a,
shahmansouri2017a,moradi2017,shahmansouri2017b,moghadam2017,shahmansouri2017c,tyshetskiy2013,
zhu2013b}, the  external magnetic field \cite{niknam2013,moradi2016b,misra2007}, the particle-particle collisions   \cite{niknam2013}, the relativistic factor \cite{zhu2013}, the particle spins \cite{shahmansouri2017a,shahmansouri2017c}, nonlocality   \cite{moradi2015,moradi2016a,moradi2017},  as well as, the effects of  exchange-correlation of plasma particles \cite{shahmansouri2015,moradi2017,shahmansouri2017b,moghadam2017,shahmansouri2017c}.   On the other hand, some attention has also been  paid to investigate  nonlinear effects in surface plasma waves. For example, Stenflo \cite{stenflo1996}   showed that   surface plasma solitary waves can appear in the vicinity of the interface between  a  plasma and the
bounding medium. 
 However, to the best of our knowledge, the theory of surface plasmons in massless Dirac plasmas has not yet been explored, and so is the subject of the present study. 
\par
In this letter, we show that the surface plasmons in massless Dirac plasmas and Fermi plasmas have several striking differences including the fact that the long-wavelength surface plasmon frequency     in massless Dirac plasmas is explicitly  nonclassical, whereas that in Fermi plasmas corresponds to the classical plasma frequency. The outline of this paper is as follows: An introduction is given in Sec. \ref{intro}. The fundamental set of dynamical equations for massless Dirac plasmas and Fermi plasmas are presented in Sec. \ref{model-equations}. Then, the dispersion relation of surface plasma waves is obtained in Sec. \ref{disp-relation}. Finally, Sec. \ref{conclusion} is left to conclude our results.
 \section{Hydrodynamic model for a massless Dirac plasma}\label{model-equations}  
 We consider the propagation of surface plasma oscillations in semi-bounded massless Dirac plasmas and Fermi plasmas. To this end, we employ the quantum hydrodynamic model applicable for   both   Dirac and Fermi plasmas with ions forming only the neutralizing background. In the fluid equations, the appropriate pressure laws for the Dirac and Fermi fluids (to be denoted, respectively, with the subscripts `D' and `F') may be discussed.  First of all, the assumption of a well-defined Fermi wavenumber $k_F$ can be valid with  the definition of the $d$-dimensional electronic density  \cite{sachdeva2015} $n_d=gk_F/2^d\pi^{d/2}\Gamma\left(1+d/2\right)$, where $g$, $d$, and $\Gamma$   are, respectively, the degeneracy factor $(g=g_sg_v$ with $g_s=2$ being the spin degeneracy and $g_v$  the pseudo-spin degeneracy factor which for graphene is $\sim2$), the system dimensionality and the Gamma function.  We, however, consider a degenerate plasma at zero temperature  in which  the energy density can be obtained as $\varepsilon=\int_0^{k_F} E(k)d^dk$, where the energy dispersion relation $E(k)$ is  expressed differently in each plasma system, given by, $\varepsilon_D=\hbar k v_F$  and $\varepsilon_F=\hbar^2 k^2/2m$. In the case of a massive Dirac fluid we have $\varepsilon\sim\hbar\sqrt{k^2+\left(\Delta/\hbar v_F\right)^2}v_F$, where $2\Delta$ is the energy gap. However, this is not  the case in our present theory. Next,  the thermodynamical identity $P=n\partial \varepsilon/\partial n-n$  can be employed to obtain the following  expressions of pressure for  the   Dirac and Fermi fluids    in  three-dimensional plasmas \cite{moghanjough2013b}
\begin{equation}
P_D=\frac{\left(3\pi^2\right)^{4/3}}{12\pi^2}v_F\hbar n^{4/3},~~  
P_F=\frac{\left(3\pi^2\right)^{2/3}}{5m_e}\hbar^2 n^{5/3}. \label{pressure-law}
\end{equation}                                                                                                                                                                                                                                                     We emphasize that  the density dependencies of  $P_D$ and $P_F$ are  different.  We also note that the QHD model can be employed for both the cases of non-relativistic quantum Fermi fluids and relativistic massless Dirac fluids. Furthermore, the QHD model for   Dirac fluids is independent of the electron mass \cite{zhu2010,mendoza2013} for which the basic equations   read
\begin{equation}
\frac{\partial n}{\partial t}+\nabla\cdot(n\mathbf{u})=0, \label{cont-eq}
\end{equation}
\begin{equation}
\begin{split}
\left(P+\varepsilon\right)\left(\frac{\partial }{\partial t}+\mathbf{u}\cdot\nabla\right)\mathbf{u}=&enc^2\left[\nabla\phi+\pmb{\beta}\left(\pmb{\beta}\cdot\nabla\right)\phi\right]  \\
&-\frac{c^2}{\gamma^2}\left(\nabla P+\frac{\pmb{\beta}}{c}\frac{\partial P}{\partial t}\right), \label{moment-eq}
\end{split}
\end{equation}
\begin{equation}
\nabla^2\phi=4\pi e\left(n-n_0\right), \label{poiss-eq}
\end{equation}                                                                                                                                                             where $n$ and $\mathbf{u}$, respectively, denote the number density and velocity of electrons, $\phi$  is the electrostatic potential,  $n_0$ is  the equilibrium  number density of electrons and ions, and $P$ is the fluid pressure. Also, $\pmb{\beta}=\mathbf{u}/c$ with $c$ denoting the speed of light in vacuum and $\gamma=1/\sqrt{1-\beta^2}$ is the relativistic factor. 
\par
In the weak relativistic limit  $P\ll\varepsilon=mnc^2$, the   pressure in Eq. \eqref{moment-eq} can be  due to the Fermi degeneracy pressure $P_F$ \cite{shahmansouri2015}. Furthermore, in unmagnetized plasmas and with $u\leq v_F\ll c$ for which $\gamma\sim1$, the following equations can be obtained for Fermi plasmas.
\begin{equation}
\frac{\partial n}{\partial t}+\nabla\cdot(n\mathbf{u})=0, \label{cont-eq-fermi}
\end{equation}
\begin{equation}
\left(\frac{\partial }{\partial t}+\mathbf{u}\cdot\nabla\right)\mathbf{u}=\frac{e}{m}\nabla\phi-\frac{1}{mn}\nabla P_F , \label{moment-eq-fermi}
\end{equation}
\begin{equation}
\nabla^2\phi=4\pi e\left(n-n_0\right), \label{poiss-eq-fermi}
\end{equation}
where $P_F$ is the Fermi pressure given by Eq. \eqref{pressure-law}, and we have neglected the quantum dispersion effect associated with the Bohm potential for simplicity and also for smallness compared to the degeneracy pressure gradient (e.g., in solid density plasmas).  
On the other hand, in  Dirac plasmas, since the Fermi speed   $v_F\sim c/300$, the  weakly relativistic condition $(\beta\ll1)$ can be employed, however, due to the different energy dispersion $E$ for the Dirac fermions and the ordinary fermions (viz., $E\sim\hbar k v_F$  and $E\sim\hbar^2 k^2 /2m$ respectively), the weak relativistic assumption does not apply to the massless Dirac fluids, and in this case, the corresponding equations read \cite{moghanjough2013b,mendoza2013}    
\begin{equation}
\frac{\partial n}{\partial t}+\nabla\cdot(n\mathbf{u})=0, \label{cont-eq-dirac}
\end{equation}
 \begin{equation}
\left(P+\varepsilon\right)\left(\frac{\partial }{\partial t}+\mathbf{u}\cdot\nabla\right)\mathbf{u}=enc^2\nabla\phi-c^2\nabla P_D, \label{moment-eq-dirac}
\end{equation}
\begin{equation}
\nabla^2\phi=4\pi e\left(n-n_0\right), \label{poiss-eq-dirac}
\end{equation}
where the physical variables $n,~\phi,~\mathbf{u}$ etc. all are functions of $\mathbf{R}$ and $t$ with $\mathbf{R}=(\mathbf{r},x)$ and $\mathbf{r}=(y,z)$.  The pressure $P_D$ in Eq. \eqref{moment-eq-dirac} represents the quantum fluid pressure for massless Dirac Plasmas given by Eq. \eqref{pressure-law}.  
In what follows, we study the basic features of surface plasma oscillations at the interface of  a massless Dirac plasma (e.g., graphene) and air. The theory of surface plasmon excitation in Fermi plasmas is well-known and has been studied extensively \cite{lazar2007,shahmansouri2015,moradi2015,shahmansouri2017a,shahmansouri2017b,moghadam2017,shahmansouri2017c}, however, we review it for Fermi plasmas and compare with that in massless Dirac plasmas. 
 \section{Dispersion relation of surface plasmons} \label{disp-relation}
In order to obtain the dispersion relation for surface plasmons in a massless Dirac plasma half-space (occupying the region  $x<0$) bounded by air ($x>0$ ),  we linearize the relevant physical quantities about their unperturbed (with suffix $0$) and perturbed (with suffix $1$)  values by letting  $n=n_0+n_1,~\mathbf{u}=\mathbf{u}_1$,  and $\phi=\phi_1$, where  $n_1\ll n_0$. Then applying the space-time Fourier transform formula  of an arbitrary function $f({\bf R},t)$, given by
\begin{equation}
f({\bf R},t)=\frac{1}{(2\pi)^3}\int\int d^3k d\omega F({\bf k},\omega;x)e^{i{\bf k}\cdot {\bf r}-i\omega t}, 
\end{equation}                                                                                            
 where ${\bf k}=\left(k_y,k_z\right)$,  to the linearized basic equations of Eqs. \eqref{cont-eq-dirac} to \eqref{poiss-eq-dirac}, we obtain
 \begin{equation}
 \frac{d^2N_1(x)}{dx^2}-\gamma_j^2N_1(x)=0, \label{N1-eq}
\end{equation}  
\begin{equation}
\frac{d^2\Phi_1(x)}{dx^2}-k^2\Phi_1(x)=4\pi eN_1(x), \label{Phi1-eq}
\end{equation}                                                                                                                         
where $N_1$ and $\Phi_1$ denote the Fourier transformed variables corresponding to $n_1$ and $\phi_1$ respectively. Furthermore, $\gamma_D=\sqrt{k^2+(\omega^2_{pD}-\omega^2)/\beta_D^2}$ (for $j=D$) is the decay variable of the wave into Dirac plasmas   with $\beta_D=c/\sqrt{3}$, and $\omega_{pD}~\left(=2\sqrt{r_s}\beta_Dn_0^{1/3}\propto1/\sqrt{\hbar}\right)$ denoting the Dirac plasma oscillation frequency. Here, $r_s=e^2/\hbar v_F$ is the quantum coupling parameter (or fine structure constant), which determines the validity of the QHD model for quantum plasmas. The values  with $r_s\leq0.1$ and $0.1<r_s\leq1$     correspond  to the weak    and moderate coupling respectively \cite{bonitz2013}. On the other hand, in  Fermi plasmas,  the corresponding wavenumber and the plasma frequency are, respectively, given by $\gamma_F=\sqrt{k^2+(\omega^2_{pF}-\omega^2)/v_{F}^2}$  and  $\omega_{pF}=\sqrt{4\pi n_0 e^2/m}$ with $v_{F}=(3\pi^2)^{1/3}\hbar n_0^{1/3}/\sqrt{3}m$. Comparing the plasma frequencies $\omega_{pD}$ and $\omega_{pF}$, we find that   $\omega_{pF}$    is exactly the same as the classical plasma oscillation frequency $\omega_{p}$), however, $\hbar$ appears explicitly in $\omega_{pD}$, i.e., $\omega_{pD}\propto1/\sqrt{\hbar}$, implying that the plasmon frequency in Dirac plasmas is non longer classical.  Furthermore, the density dependency of the plasma frequency in massless Dirac plasmas is different  from that in   Fermi plasmas, i.e.,  $\omega_{pD}\sim n_0^{1/3}$  and $\omega_{pF}\sim n_0^{1/2}$.
\par 
Out of several possible plasma modes \cite{misra2011,misra2007,misra2009}, we are interested in the solutions of Eqs.  \eqref{N1-eq} and \eqref{Phi1-eq} that have the following forms 
 \begin{equation}
 N_1(x)=\Big\lbrace\begin{array}{cc}
 0 & x\geq0\\
 A\exp(\gamma_j x) & x\leq0 
  \end{array}
\end{equation}                                                                                                                 
and 
   \begin{equation}
 \Phi_1(x)=\Big\lbrace\begin{array}{cc}
 A\exp(-k x) & x\geq0\\
 C\exp(k x)+D\exp(\gamma_j x) & x\leq0  
  \end{array}
\end{equation}                                                                                                                                                                                                               
where  $A,~B,~C$ and $D$ are unknown constant coefficients to be determined   by using the following boundary conditions.
\begin{equation}
\Phi_{\text{in}}(x)\Big|_{x=0^-}=\Phi_{\text{out}}(x)\Big|_{x=0^+}, 
\end{equation} 
\begin{equation}
\frac{\partial\Phi_{\text{in}}(x)}{\partial x}\Big|_{x=0^-}=\frac{\partial\Phi_{\text{out}}(x)}{\partial x}\Big|_{x=0^+},
\end{equation} 
\begin{equation}
\frac{4P_{D0}}{3n_0}\frac{\partial N_{1}(x)}{\partial x}\Big|_{x=0^-}=en_0\frac{\partial\Phi_{\text{in}}(x)}{\partial x}\Big|_{x=0^-},
\end{equation} 
 where the subscripts \textit{in} and \textit{out} refer,  respectively,  to inside and outside   the plasma half-space, and $P_{D0}$ is the value of $P_D$ at $n=n_0$. It must be emphasized that the surface waves restrict to that part of solutions which decay away from the interface in both the regions. 
Then, after some algebra, we obtain the following dispersion  relation for surface plasmons.
\begin{equation}
2\beta_D^2\gamma_D\left(\gamma_D+k\right)=\omega_{pD}^2. \label{dispersion-dirac}
\end{equation}                                                                                                                                                                                                                                                   The corresponding dispersion relation in Fermi plasmas   can be obtained by replacing  $\gamma_D\rightarrow\gamma_F,~\omega_{pD}\rightarrow\omega_{p}$ and $\beta_D\rightarrow\sqrt{5/3}v_{F}$ as
\begin{equation}
2v_{F}^2\gamma_F\left(\gamma_F+k\right)=\omega_{p}^2. \label{dispersion-fermi}
\end{equation}                                                                                                                                                                                                                                                  
In the overcritical density plasma limit, i.e.,  $k^2\beta_D^2\ll |\omega_{pD}^2-\omega^2|$, Eqs.   \eqref{dispersion-dirac} and \eqref{dispersion-fermi} give  the following   surface modes in massless Dirac plasmas and Fermi plasmas.
 \begin{equation}
 \omega_D\approx \frac{\omega_{pD}}{\sqrt{2}}\left(1+\frac{k\beta_D}{\sqrt{2}\omega_{pD}}\right),\label{dispersion-dirac-reduced}
 \end{equation}
 \begin{equation}
 \omega_F\approx \frac{\omega_{p}}{\sqrt{2}}\left(1+\frac{kv_{F}}{\sqrt{2}\omega_{p}}\right). \label{dispersion-fermi-reduced}
 \end{equation} 
Equation \eqref{dispersion-dirac-reduced} [\eqref{dispersion-fermi-reduced}] describes the frequency of surface plasma waves that may propagate in a Dirac [Fermi] plasma half-space bounded by air.  It is to be  mentioned that we have derived the dispersion relations \eqref{dispersion-dirac} and \eqref{dispersion-fermi}   on the assumption that the interface between the plasma and air   is sharp which is valid when the surface skin depth is much larger than   the width of the physical transition layer between the two media. However, when this condition is relaxed the dispersion relations \eqref{dispersion-dirac-reduced} and \eqref{dispersion-fermi-reduced}  should be modified. It has been shown in Ref. \cite{marklund2008} that  when there is a transition layer of finite width,  the effects of  quantum broadening of the transition layer can lead to wave damping in the propagation of surface plasmon polaritons in quantum plasmas (Eq. (9) in Ref. \cite{marklund2008}). In particular,  Eq. (9) in Ref. \cite{marklund2008} recovers    the same frequency as  $\omega_F$ in Eq. \eqref{dispersion-fermi-reduced} when one ignores the   imaginary part (due to the sharp boundary),  adjusts with the dielectric constant, and replaces the small factor $0.6$ by $0.5$.        
 Next, we note that though  the forms of the dispersion relations \eqref{dispersion-dirac-reduced} and \eqref{dispersion-fermi-reduced}   are apparently similar, however,  their properties are qualitatively different due to the appearance of $\omega_{pD}$ and $\beta_D$ in $\omega_D$, while $\omega_{p}$ and $v_F$ in $\omega_F$. 
\par 
A comparison between $\omega_D$  and $\omega_F$ may be made.      In the long-wavelength limit $K\equiv k\beta_D/\omega_{pD}\ll1$, we have $\omega_F\sim\omega_p$, i.e., the usual plasma oscillation frequency and so the wave frequency becomes classical in nature, however,    $\omega_D\sim\omega_{pD}/\sqrt{2}\propto1/\sqrt{\hbar}$,     implying that  the surface plasmon frequency in massless Dirac plasmas is explicitly  non-classical and does not have a classical plasma frequency analogy. Furthermore, for a finite  $K$,  $\omega_F$ is essentially quantum mechanical in nature having dependency on  $v_F\propto1/\hbar$.
Besides the non-classical nature of $\omega_D$, there are also other consequences that separates $\omega_D$ from $\omega_F$. For example, the density dependency of $\omega_D$  is clearly different from  $\omega_F$, namely    $n^{1/3}$ and $n^{1/2}$ respectively.   The other difference between  $\omega_D$  and $\omega_F$ is the  appearance of the fine structure constant $r_s$ in the cutoff frequency of $\omega_D$    viz.,  $\omega_D\Big |_{K\rightarrow0}\propto r_s^{1/2}$, however, not present in $\omega_F$.   
\section{Conclusion}\label{conclusion}
We have studied the existence and propagation characteristics  of surface plasma waves in a semi-bounded (bounded by air) massless Dirac plasma (such as those occurring in doped graphene layers and interacting through the long-range Coulomb force).  A quantum hydrodynamic model is used to derive the dispersion relation for these surface waves. 
It is found that the surface plasmon mode in a massless Dirac plasma  has several striking differences compared to that in a Fermi plasma. For example, in the long-wavelength limit (i.e., $k\beta_D/\omega_{pD}\ll1$),  the surface plasma waves in  massless Dirac plasmas propagate below the Dirac-plasma frequency, i.e.,  $\omega_D~\sim\omega_{pD}/\sqrt{2}$,  and  is  explicitly non-classical, its frequency being proportional to $1/\sqrt{h}$. The origin of such non-classical wave mode is the consequence of the  relativistic Dirac quantum electron fluids.  This  is, however, in contrast to the case of Fermi plasmas where $\omega_F~\sim\omega_{p}/\sqrt{2}$, i.e., the wave frequency is truly classical in the long-wavelength limit. 
In the latter, the density dependencies of the wave frequencies   in both the cases are also different, i.e., $\omega_D\propto n_0^{1/3}$ and $\omega_F\propto n_0^{1/2}$.  
Another striking difference is the appearance of the quantum coupling parameter (fine structure constant) $r_s$ in the surface wave frequency $\omega_D$ which is also not present in the case of Fermi plasmas. 
 \par 
To conclude,  the results of the present study can be useful for    understanding the salient features of surface plasma waves that can be excited in semi-bounded massless Dirac plasmas,  such as those in graphene superlattices or nanoribbons.


\section*{Acknowledgement} {One of us (A.P.M) acknowledges  support from UGC-SAP (DRS, Phase III) with Sanction  order No.  F.510/3/DRS-III/2015(SAPI),  and UGC-MRP with F. No.
43-539/2014 (SR) and FD Diary No. 3668.}


\begin{thebibliography}{00}
\bibitem{pines1966}   D. Pines, P. Nozieres, The Theory of Quantum Liquids (W.A. Benjamin, Inc., New York, 1966). 
 \bibitem{giuliani2005} G.F. Giuliani, G. Vignale, Quantum Theory of the Electron Liquid (Cambridge University Press, Cambridge, 2005). 
 \bibitem{bonaccorso2010} F. Bonaccorso, Z. Sun, T. Hasan, A.C. Ferrari, Nature Photon 4  (2010)  611. 
 \bibitem{geim2011} A.K. Geim, Rev. Mod. Phys. 83  (2011) 851. 
 \bibitem{novoselov2004} K.S. Novoselov, A.K. Geim, S.V. Morozov, D.J. Iang, Y.Z. Hang, S.V. Dubonos, I.V. Grigorieva,  A.A. Firsov, Science 306   (2004) 666. 
\bibitem{grigorenko2012} A.N. Grigorenko, M. Polini,   K.S. Novoselov, Nature Photon. 6 (2012) 749.
 \bibitem{fei2013} Z. Fei, A.S. Rodin, W. Gannett, S. Dai, W. Regan, M. Wagner, M.K. Liu, A.S. McLeod, G. Dominguez, M. Thiemens {it et al.}, Nat. Nanotechnol. 8   (2013) 821–825. 
 \bibitem{li2008} Z.Q. Li, E.A. Henrikse, Z. Jiang, Z. Hao, M.C. Martin, P. Kim, H.L. Stormer,   D.N. Basov, Nat. Phys. 4  (2008)  532–535. 
 \bibitem{neto2009} A.H.C. Neto, F. Guinea, N.M.R. Peres, K.S. Novoselov,  A.K. Geim, Rev. Mod. Phys. 81   (2009) 109. 
\bibitem{liu2014} S. Liu, C. Zhang, M. Hu, X. Chen, P. Zhang, S. Gong, T. Zhao,  R. Zhong,  Appl. Phys. Lett. 104  (2014)  201104. 
\bibitem{bonaccorso2012} F. Bonaccorso, A. Lombardo, T. Hasan, Z. Sun, L. Colombo,  A.C. Ferrari, Mater. Today 15   (2012) 564. 
\bibitem{peres2010} N.M.R. Peres, Rev. Mod. Phys. 82   (2010) 2673. 
\bibitem{ando1982} T. Ando, A.B. Fowler,  F. Stern, Rev. Mod. Phys. 54  (1982) 437. 
\bibitem{mahan2000} G.D. Mahan, Many Particle Physics (Plenum Publisher, New York, 2000). 
\bibitem{li1991} Q.P. Li, S. Das Sarma, Phys. Rev. B 43  (1991) 11768 . 
 \bibitem{moghanjough2013a} M. Akbari—Moghanjough, Phys. Plasmas 20   (2013) 102115. 
 \bibitem{sarma2009} S. Das Sarama, E.H. Hwang, Phys. Rev. Lett 102  (2009)  206412. 
\bibitem{sachdeva2015} R. Sachdeva, A. Thakur, G. Vignale,   A. Agrawal, Phys. Rev B 91  (2015)  205426.
\bibitem{ju2011} L. Ju, B. Geng, J. Horng, C. Girit, M. Martin, Z. Hao, H.A. Bechtel, X. Liang, A. Zettl, Y. Ron Shen, F. Wang, Nat. Nanotechnol. 6  (2011)  630. 
\bibitem{yan2012} H. Yan, X. Li, B. Chandra, G. Tulevski, Y. Wu, M. Freitag, W. Zhu, P. Avouris, F. Xia, Nat. Nanotech. 7 (2012)  330 . 
\bibitem{yoon2014} H. Yoon, C. Forsythe, L. Wang, N. Tombros, K. Watanabe, T. Taniguchi, J. Hone, P. Kim, D. Ham, Nat. Nanotech. 9  (2014)  594. 
\bibitem{yan2013} H. Yan, T. Low, W. Zhu, Y. Wu, M. Freitag, X. Li, F. Guinea, P. Avouris, F. Xia, Nat. Photonics 7   (2013) 394. 
\bibitem{stauber2014} T. Stauber, J. Phys.: Condens. Matter 26  (2014) 123201.
\bibitem{pietro2013} P. Di Pietro, M. Ortolani, O. Limaj, A. Di Gaspare, V. Giliberti, F. Giorgianni, M. Brahlek, N. Bansal, N. Koirala, S. Oh, P. Calvani, S. Lupi , Nat. Nanotech. 8   (2013) 556. 
\bibitem{raghu2010} S. Raghu, S.B. Chung, X.-L Qi, and S.-C. Zhang, Phys. Rev. Lett. 104   (2010) 116401. 
\bibitem{wunsch2006} B. Wunsch, T. Stauber, F. Sols,   F. Guinea, New J. Phys.  8  (2006) 318. 
\bibitem{hwang2007} E.H. Wang, S. Das Sarma, Phys. Rev. B 75   (2007) 205418. 
\bibitem{polini2008} M. Polini, R. Asgari, G. Borghi, Y. Barlas, T. Pereg-Barnea,   A.H. MacDonald, Phys. Rev. B 77  (2008) 081411(R).
\bibitem{abedinpour2011} S.H. Abedinpour, G. Vignale, A. Principi, M. Polini, W.K. Tse,  A.H. MacDonald, Phys. Rev. B 84   (2011) 045429.
\bibitem{zhou2015} J. Zhou, H.-R. Chang, D. Xiao, Phys. Rev. B 91   (2015) 035114.
\bibitem{zhu2013} J.-J. Zhu, S.M. Badalyan,   F.M. Peeters, Phys. Rev. B 87 (2013)  085401.
\bibitem{triola2012} C. Triola, E. Rossi, Phys. Rev. B 86   (2012) 161408 (R).
\bibitem{ritchie1963} R.H. Ritchie, Prog. Theor. Phys. 29   (1963) 607.
\bibitem{lazar2007} M. Lazar, P.K. Shukla,   A. Smolyakov, Phys. Plasmas 14  (2007) 124501. 
 \bibitem{chang2008} I-S Chang, Y-D Jung, Phys. Lett. A 372 (2008) 1498.
\bibitem{stenflo1996} L. Stenflo, Phys. Scr. T63 (1996) 59.
\bibitem{kaw1970} P.K. Kaw, J.B. McBride, Phys. Fluids 13   (1970) 1784.
\bibitem{misra2011} A.P. Misra, Phys. Rev. E 83   (2011) 057401.
\bibitem{misra2010} A.P. Misra, N.K. Ghosh,   P.K. Shukla, J. Plasma Phys. 76  (2010)  87.
\bibitem{niknam2013} A.R. Niknam, S.T. Boroujeni,   S.M. Khorashadizadeh, Phys.Plasmas 20   (2013) 122106.
\bibitem{shahmansouri2015} M. Shahmansouri, Phys. Plasmas 22  (2015)  092106.
\bibitem{moradi2015} A. Moradi, Phys. Scr. 22   (2015) 014501.
\bibitem{moradi2016a} A. Moradi, Phys. Plasmas 23  (2016)  084501.
\bibitem{shahmansouri2017a} M. Shahmansouri, B. Faokhi, R. Aboltaman, Phys. Plasmas 24 (2017) 054505. 
\bibitem{moradi2017} A. Moradi, Comm. Theor. Phys. 67  (2017)  317. 
\bibitem{shahmansouri2017b} M. Shahmansouri, M. Mahmodi Moghadam, Phys. Plasmas 24  (2017)  102107.
\bibitem{moghadam2017} M. Mahmodi Moghadam, M. Shahmansouri,   B. Farokhi, Phys. Plasmas 24   (2017) 122102. 
\bibitem{shahmansouri2017c} M. Shahmansouri, A.P. Misra, Plasma Sci. Tech., in press (2017). 
\bibitem{tyshetskiy2013} Y.O. Tyshetskiy, S.V. Vladimirov,   R. Kompaneets, J. Plasma Phys.79   (2013) 387. 
\bibitem{zhu2013b} J. Zhu, H. Zhao,   M. Qiu, Phys. Lett. A 377  (2013)  1736. 
\bibitem{moradi2016b} A. Moradi, Phys. Plasmas 23  (2016) 044701. 
\bibitem{misra2007} A.P. Misra, Phys. Plasmas 14   (2007) 064501. 
\bibitem{moghanjough2013b} M. Akbari Moghanjough, J. Appl. Phys. 114   (2013) 073302. 
\bibitem{zhu2010} J. Zhu and P. Ji, Phys. Rev. E 81  (2010)  036406.  
 \bibitem{mendoza2013} M. Mendoza, H.J. Herrmann,  S. Succi, Sci. Rep. 3   (2013) 1052. 
\bibitem{bonitz2013} M. Bonitz, E. Pehlke,  T. Schoof, Phys. Rev. E 87   (2013) 033105.
\bibitem{misra2009} A.P. Misra, Phys. Plasmas 16   (2009) 074505.
\bibitem{marklund2008} M. Marklund, G. Brodin, L. Stenflo, C.S. Liu, Europhys. Lett.  84 (2008) 17006.
\end{thebibliography}
\end{document}